# An Intrinsic Advantage of Sexual Reproduction

Xiang-Ping Jia[*] and Hong Sun[§]

[*]*College of Science, Liaoning University of Technology, Jinzhou, P. R. China 121001*

[§]*College of Foreign Languages, Liaoning University of Technology, Jinzhou, P. R. China 121001*

E-mail: *jiaxiangping@tsinghua.org.cn*

---

[1] Xiang-Ping Jia Shiying 169 Jinzhou Liaoning China 121001

## Abstract

The prevalence of sexual reproduction ("sex") in eukaryotes is an enigma of evolutionary biology. Sex increases genetic variation only tells its long-term superiority in essence. The accumulation of harmful mutations causes an immediate and ubiquitous pressure for organisms. Contrary to the common sense, our theoretical model suggests that reproductive rate can influence the accumulation of harmful mutations. The interaction of reproductive rate and the integrated harm of mutations causes a critical reproductive rate $R^*$. A population will become irreversibly extinct once the reproductive rate reduces to lower than $R^*$. A sexual population has a $R^*$ lower than 1 and an asexual population has a $R^*$ higher than 1. The mean reproductive rate of a population which has reached to the carrying capacity has to reduce to 1. That explains the widespread sex as well as the persistence of facultative and asexual organisms. Computer simulations support significantly our conclusion.

# Introduction

Unlike simple and efficient asexual reproduction, sexual reproduction ("sex") has many disadvantages. There is a twofold cost for males with sex compared with asexual organisms (Williams 1975). Meiosis and syngamy spend energy and resources not needed for asexual reproduction (Crow 1994; Maynard Smith 1978). Genetic segregation and recombination often broke up favourable genotypes. Sex transfers diseases and harmful transposons. Sexual selection often leads to maladaptive traits, such as the peacock's tail. These and more additional defects imply sex has a more than twofold cost as asexual reproduction. As compensation, sex may be a more efficient means for transferring genes to further generations. However, a sexual parent transfers only 50% of its genes to the next generation compared with 100% for an asexual parent (Otto 2009). For all this, sex has spread to almost all eukaryotic organisms. Why did this reproductive approach overcome its many defects to persist so widely? This has been an important issue in evolutionary thoughts since Darwin published his epochal evolutionary theory (Darwin 1859). The solution accounting for the ubiquity of sex should give the long-term and short-term advantages of sex because more efficient asexual individuals might invade a sexual population.

***Increasing Genetic Variation.*** Weismann was the first to state that sex causes genetic variation which increases the rate of evolution of populations (Weismann 1889). Although that sex evolved to cause variation may well be correct, but there are two holes with it (Otto 2009). In the case of a single gene subject to selection, in which homozygotes have the high fitness relative to heterozygotes, to reproduce asexually will preserve more genetic variations than to reproduce sexually in a population. Also, in a population with heterozygous advantage to reproduce sexually causes variation but reduces fitness. To clarify this issue, we need to confirm how sex or asexual reproduction affects on the genetic variation of a population? And what is the advantage of genetic variation in the face of nature selection?

Genetic variation implies gene diversity between individuals in a population. We define *life cycle of a mutation* as the time from its arising to being removed or fixed by selection or random drift in a population. Assume

$n_m$ is the mean life cycle (generations) of mutations in a population,

$U$ is the mutation rate, that the number of new mutations with an offspring and

$n$ is the number of generations back needed to reach the shared ancestor of two individuals.

Therefore, the mean number of new mutations with each individual is $Un_m$. The assumption is that a mutation never arises synchronously and spontaneously with two separated individuals. In an asexual population, the number of different mutations between two individuals is

$$M_{Asex} = Un \tag{1}$$

The gene diversity between individuals is linearly related to $n$ in an asexual population.

We can ignore the new mutations' contribution to genetic variation for simplicity, if a sexual individual has far more mutations than the new mutations in each generation ($Un_m >> U$). An individual inherits $Un_m/2^n$ mutations from an ancestor $n$ generations ago in a sexual population. The number of different mutations between two individuals with a shared ancestor $n$ generations ago is

$$M_{Sex} = (1 - 1/2^{2n})Un_m \qquad (2)$$

If these two sexual individuals have two shared ancestors, the right hand of Eq. 2 becomes $(1 - 1/2^{2n_1} - 1/2^{2n_2})Un_m$, where $n_1$ and $n_2$ are the number of generations back to these two ancestors. Figure 1 shows Eq. 1 with a dashed line and Eq. 2 with a solid line. Both sex and asexuality can cause genetic variation, but sex causes variation more quickly than asexuality. This might have answered why sex is so widespread, if genetic variation causes more of fitter individuals in evolution.

The mean life cycle of mutations $n_m$ is different for each population. It is related to the population size and the mean reproductive rate. More favourable mutations also results in a greater $n_m$. Besides, $n_m$ will increase in a sexual population with heterozygous advantage.

More genetic variations imply that the relative fitness of individuals distributes more evenly in a population. Figure 2A shows the relative fitness distributions of populations with more and few genetic variations. In addition, the proportion of individuals eliminated in a population can tell the strength of selection. Strong selection will eliminate more individuals than the weak selection (Fig. 2B and 2C).

Few individuals survive after a strong selection. The surviving proportion of a population with more genetic variations is larger than which with few genetic variations (Fig. 2C). Namely, a population with few genetic variations is more likely to be extinct under strong selection. In contrast, weak selection will eliminate few individuals. The eliminated proportion of a population with more genetic variations is larger than that with few genetic variations (Fig. 2B). A population with more genetic variations needs a heavier reproduction load than one which with few genetic variations under weak selection. Sex causes more genetic variations so it is favourable when selection is strong. However, asexual reproduction is favourable under weak selection because of the inherent more than twofold advantage and the lighter reproduction load.

Strong selection arises occasionally and weak selection plays a leading role at most of the time in real populations. By reasoning, more genetic variations bring species a long-term advantage to pass catastrophes that imply strong selection. Facultative reproduction may be the best because of the alternation of strong and weak selection. Thus, genetic variation is not enough to explain why most eukaryotes reproduce obligately sexually.

These are examples of facultative organisms. (i) The most common of vegetative growth in yeast is asexual reproduction by budding (Balasubramanian *et al.* 2004; Yeong 2005). Under high stress conditions such as starvation, diploid cells can undergo sporulation, entering sexual reproduction and producing haploid spores (Neiman 2005). (ii) Aphids produce eggs parthenogenetically without meiosis in spring and summer (Blackman 1979; Hales *et al.* 2002). In late autumn as days become shorter and temperatures fall, aphids begin producing eggs sexually on perennial host plants to pass the winter. These two organisms reproduce asexually in comfortable condition (weak selection) and sexually in severe conditions (strong selection), so they engage in facultative reproduction instead of obligately sexual or asexual reproduction.

*Faster Adaptation to Changing Environment.* An environment often changes because of migrations or changing natural conditions as temperature, humidity and resource supply. More different offspring will increase adaption to changing environments and sex is more efficient than asexual reproduction to cause more genetic variations in offspring. For example, the lottery model relies on the area of occurrence of any species that is mostly heterogeneous to a greater or lesser degree (Williams 1975). A polymorphic sexual population is more adaptive to different microhabitats than an asexual population. The tangled bank hypothesis (Bell 1982) declares that in enough complicated ecosystems, lines and species that reproduce sexually have a greater chance of survival in the long run. Based on the experiments on rotifers showing facultative sex (Becks and Agrawal 2010; Becks and Agrawal 2012), Roze (Roze 2012) supposed the idea that rapid adaptation to changing environments relies on the benefits of increasing genetic variation.

These theories depend essentially on sex increasing genetic variation. Because causes genetic variation, sex is logically favourable if the environment changes always severely enough and continually, that is under strong selection. However, catastrophes are scarce and there are only mild changes at most of the time in real environments. In addition, organisms migrate with the purpose to look for a more suitable environment to survive, so migrations weaken the environment's changes.

Therefore, sex brings about a population a long-term but not a short-term advantage to adapt to a changing environment.

*Competition between Species.* The competition between antagonistic species, as hosts and parasites (Hamilton 1980; Jaenike 1978; Jokela *et al.* 2009; Lloyd 1980), induces a continuous selection pressure. This is the famous *Red Queen hypothesis* for the evolution of sex (Van Valen 1973). It is effective in many experiments on clonal fish (Lively *et al.* 1990), snails (Jokela *et al.* 2009; Lively 1987; Lively and Jokela 2002) and psychid moths (Kumpulainen *et al.* 2004), but is not obvious in some experiments on Poecilia (Tobler and Schlupp 2005). Hamilton et al (Hamilton *et al.* 1990) proposed the idea of parasite coevolution with sex was better than the previous models, such as very low fecundity, realistic patterns of genotype fitness and changing environments.

Regrettably, in some cases as the rapidly disequilibria in evolution of haploid models, the Red Queen hypothesis requires that selection must be quite strong (Howard and Lively 1994; May and Anderson 1983; Otto and Nuismer 2004). Selection does actually not necessarily have to be strong in both species. Strong selection in the parasites and moderately weak selection in the host favour the evolution of sex (Salathe et al. 2008). Fatal infection does not often occur in hosts; parasites have only weak virulence and infect only those individuals with low fitness at most of the time. In essence, Red Queen hypothesis depends also on genetic variation and not enough to explain why most of eukaryotes reproduce obligately sexually rather than facultatively.

***Bringing Favourable Mutations Together.*** The works by Fisher and Muller (Fisher 1930; Muller 1932) showed that recombination can bring together favourable mutations arising in different individuals in the same individual to evolve in parallel. There may be some individuals combining favourable mutations from different parents. However, they are difficult to stand out under comfortable conditions in a short time so the population cannot withstand the immediate invasion of more efficient asexual individuals. Therefore, to bring favourable mutations together is a long-term but not short-term advantage for a population. Also, the "Fisher-Muller" hypothesis does not explain how or why sex creates beneficial genetic combinations more often than it destroys them (Maynard Smith 1971; Williams and Mitton 1973).

***Accumulation of Harmful Mutations.*** Many experiments have proved that most mutations are harmful (Eyre-Walker and Keightley 2007; Kondrashov 1988; Lynch et al. 1993). An asexual population may irreversibly suffer decline in mean fitness because of no recombination between individuals and lack of back mutations and favourable mutations (Felsenstein 1974; Muller 1964). If each progeny in a population suffers new mutations, mutation free individuals become more and more rare. No finite population is free of the decline in fitness thanks to the accumulation of harmful mutations. This is *Muller's ratchet* (Muller 1964). In an infinite population, the selection-mutation balance may restrain this decline in fitness (Butcher 1995). *Reducing the mutation load* (Crow 1994; Kondrashov 1988) is the other model that applies allegedly to whatever size population. The number of mutations removed with each eliminated individual is much larger in a sexual than in an asexual population, and the mutation load reduces correspondingly.

The endless accumulation of harmful mutations causes a more continuous decline in fitness than other reasons. No real population is infinite although it might be right that an infinite population can halt the ratchet. Seemingly, the ratchet effect causes an immediate evolution of sex. However, two further problems still exist. First, what is the mechanism with which the recombination of gene and sex halt the accumulation of harmful mutations and defeat asexual reproduction that is inherently superior? Second, why many organisms, such as most of protists and fungi, some plants and "ancient asexual scandals" bdelloid rotifers (Judson and Normark 1996; Normark *et al.* 2003), reproduce still asexually if no asexual species escapes from the ratchet effect? Are there other approaches halting the ratchet except for sex and recombination?

We introduce a parameter "Mutation Specificity" reflecting statistically the integrated harm of mutations within an individual. From theoretical inference, the interaction of the reproductive rate and the integrated harm of mutations causes a critical point ($R^*$). A population becomes irreversibly extinct once the mean reproductive rate is lower than $R^*$. Simulations support this conclusion as well.

## Mutation Specificity

A favourable mutation is more likely to be fixed than a harmful one in a population. The spread extent of a mutation is positively related to its favourable level in the statistical sense. In other word, the frequency of a mutation is negatively related to its harmful level. The individual with more low frequency mutations is more likely to harbour more harmful mutations.

We define Mutation Specificity (*S*) as the distinctive degree of a mutation. If a mutation belongs to only one individual then *S* = 1. If a mutation has spread to all individuals in an infinite population then *S* = 0. If a mutation has spread to *n* individuals, its Mutation Specificity is

$$S = 1/\sqrt[g]{n} \qquad (3)$$

The parameter *g*, which is equivalent to a mutation's surviving generations, influences signally Mutation Specificity. If *g* is small (as *g* = 1), the contribution of new mutations is enlarged and the contribution of mutations with high frequency is despised. We should determine a *g* greater than 1 or approaching to the mean surviving generations of mutations and then Mutation Specificity indicates better the mutations' effect on an individual.

*S* changes from 1 to 0 while fixing a mutation in a population. The Mutation Specificity of an individual, as the summation of that for all mutations, indicates the integrated harm of mutations. In general, an individual with a high Mutation Specificity harbours more of harmful mutations. Mutation Specificity is negatively related to the fitness of an individual in the statistical sense.

Assume the reproductive rate (*R*) is the number of offspring for each parent, if the offspring passed through his or her lifetime conforming to the age-specific fertility. *R* is also called net reproductive rate and is proportional to the fitness. The Mutation Specificity of a daughter in an asexual population is

$$S_{Daughter} = U + S_{Mother}/\sqrt[g]{R} \qquad (4)$$

where $U$ is the number of the daughter's new mutations and $S_{mother}/\sqrt[g]{R}$ is the Mutation Specificity inherited from her mother. An individual with more siblings inherits a lesser Mutation Specificity from her mother. If $R$ is unchanging for all ancestors, the Mutation Specificity of an asexual individual is

$$S_{Asex} = U + \frac{1}{\sqrt[g]{R}}(U + \frac{1}{\sqrt[g]{R}}(U + \cdots)) \tag{5}$$

where the first $U$ is her own new mutation, the second $U$ is her mother's, the third $U$ is her grandmother's and so on. The simplified equation is

$$S_{Asex} = U/(1 - 1/\sqrt[g]{R}) \tag{6}$$

The Mutation Specificity of an individual in an asexual population is positively related to the mutation rate $U$ and negatively related to the reproductive rate $R$ from Eq. 6.

We study only monogamy so sexual reproduction means monogamy in this study. A sexual mother needs to produce twice as many offspring as an asexual mother given the same reproductive rate. We can calculate the spreading extent of a mutation of the parents, which is inherited by offspring, by a binomial probability

$$\begin{aligned} &\frac{1}{1-2^{-n}} \left( \underbrace{C_n^1 (\frac{1}{2})^1 (1-\frac{1}{2})^{n-1} \cdot 1}_{\text{probability that one sibling inherits this mutation}} + \underbrace{C_n^2 (\frac{1}{2})^2 (1-\frac{1}{2})^{n-2} \cdot 2}_{\text{probability that two sibling inherit this mutation}} + \cdots + \underbrace{C_n^n (\frac{1}{2})^n (1-\frac{1}{2})^0 \cdot n}_{\text{probability that } n \text{ siblings inherit this mutation}} \right) \\ &= \frac{1}{1-2^{-n}} \frac{n}{2} \left( C_{n-1}^0 (\frac{1}{2})^0 (1-\frac{1}{2})^{n-1} + C_{n-1}^1 (\frac{1}{2})^1 (1-\frac{1}{2})^{n-2} + \cdots + C_{n-1}^{n-1} (\frac{1}{2})^{n-1} (1-\frac{1}{2})^0 \right) \\ &= \frac{n/2}{1-1/2^{-n}} \end{aligned} \tag{7}$$

where $n$ is the number of offspring,

"$1/2$" is the probability that an offspring inherits a certain mutation of parents,

"$1-1/2$" is the probability that an offspring misses a certain mutation of parents and

"$1-2^{-n}$" is the probability that a mutation of parents is inherited.

The number of offspring of a couple is $n = 2R$ and an offspring inherits a mutation with probability 0.5 in a sexual population, so an offspring's Mutation Specificity is

$$S_{Offspring} = U + 0.5 \cdot 2 S_{Parent} \sqrt[g]{(1-2^{-2R})/(2R/2)} = U + S_{Parent} \sqrt[g]{(1-2^{-2R})/R} \tag{8}$$

The second term of right hand of Eq. 8 is the offspring's Mutation Specificity, inherited from parents, that is $\sqrt[g]{1-2^{-2R}}$ times of that of Eq. 4. This implies that a sexual individual inherits lesser Mutation Specificity from parents than an asexual individual given the same reproductive rate. If $R$ is unchanging for all ancestors, the Mutation Specificity of a sexual individual is

$$S_{Sex} = U + \sqrt[g]{(1-2^{-2R})/R}(U + \sqrt[g]{(1-2^{-2R})/R}(U + \cdots)) \tag{9}$$

The simplified equation is

$$S_{Sex} = U \left/ \left(1 - \sqrt[g]{(1-2^{-2R})/R}\right)\right. \tag{10}$$

The Mutation Specificity of an individual in a sexual population is positively related to $U$ and negatively related to $R$ from Eq. 10. That is similar to the case of an asexual population in Eq. 6.

Equation 6 and 10 tell a novel result that the accumulation of harmful mutations within progenies depends on the reproductive rate of a lineage. It is intricate the reproductive rate influences the accumulation of harmful mutations because we commonly consider the latter determines the former. Mutation Specificity tells us that an individual inherits a favourable mutation from a distant ancestor with an increased probability compared with a harmful mutation, and reproductive rate influences signally this effect.

It is questioned that Mutations Specificity may not reflect the integrated harm of mutations because the frequency of a mutation is not negatively related to its harmful level sometimes. A harmful mutation spreads to a population with the same probability as a favourable one if natural selection does not play a role because of the very weak effect of a mutation. For example, a phenotypic variation, which causes an individual being eliminated by natural selection, may needs the integrated effect of $N_{ph}$ harmful mutations and $N_{ph}$ is larger than the mean number of mutations within each individual $Un_m$. That implies mutations have been fixed without natural selection. However, the number $N_{ph}$ reduces to $N_{ph}-x$ when $x$ harmful mutations are fixed in a population because these fixed mutations influence each individual and are not counted because of exceeding the life cycle. Natural selection begins to deal with mutations when $N_{ph}$ reduces to smaller than $Un_m$. Therefore, natural selection must play an important role on the mutations within each existing natural population so Mutation Specificity can indicate the integrated harm of mutations.

## Theoretical Result

We adopt a classical assumption which all mutations are equivalent and independent in their effect on fitness (Felsenstein 1974). Each mutation has the effect $d$ and mutations within an individual interact multiplicatively (Butcher 1995). Thus, $d > 1$ if a mutation is favourable and $d < 1$ if a mutation is harmful. Therefore, $d-1$ can reflect the favourable level of a mutation and $1/d-1$ can reflect the harmful level of a mutation. Under the multiplicative mutation interactive model, the integrated harm of many mutations for an individual $D = 1/\prod d_i - 1$.

We define the fitness $F$ of an individual as the reproductive capacity. An individual's fitness $F$ relies primarily on the mutations. An individual trends to reproduce with the highest reproductive rate, the fitness $F$, when there is not other restriction. Therefore, $F = F_0 \prod d_i$ where $F_0$ is the initial fitness and we get the equation

$$D = F_0/F - 1 \qquad (11)$$

This is shown with line $c$ (thick line) in Fig. 3.

It is easy to find that Mutation Specificity is related to the integrated harm of mutations. The simulation results (Fig. 4) supports this conclusion and even they are in direct proportion, $D = kS$ where $k$ is a constant. From Eq. 6, we get the equation for an asexual lineage.

$$D_{Asex} = kU / (1 - 1/\sqrt[g]{R}) \qquad (12)$$

From Eq. 10, we get the equation for a sexual lineage.

$$D_{Sex} = kU / \left(1 - \sqrt[g]{(1 - 2^{-2R})/R}\right) \qquad (13)$$

Equation 12 and 13 are shown with line $a$ (dashed line) and line $b$ (solid line) in Fig. 3.

The reproductive rate is the cause and the integrated harm is the effect with line $a$ and $b$ in Fig. 3. The integrated harm is the cause and the fitness is the effect with line $c$ in Fig. 3. Two kinds of lines cross at a point that gives a special reproductive rate $R^*$, such as $R*_1$ and $R*_2$ in Fig. 3. An irreversible process, in which the reducing reproductive rate and the increasing integrated harm will continually mutual promotion help each other forward, is triggered when the reproductive rate of a lineage reduces to lower than $R*$. Thereafter, the lineage becomes gradually extinct. Therefore, $R*$ is the critical reproductive rate causing a lineage extinct. This critical reproductive rate will increase with the increase of mutation rate.

An accelerated process, in which a population is irreversibly extinct because of harmful mutations, is called mutational meltdown (Allen *et al.* 2009; Gabriel *et al.* 1993; Lynch *et al.* 1993). The accumulation of harmful mutations causes the reproductive rate to reduce to lower than 1 and the population size begins to decline, initiating the meltdown (Lynch et al. 1993). That is undoubtedly correct, however, we know more from Fig. 3. First, the meltdown causing the extinction of a lineage begins if the reproductive rate reduces to lower than $R*$ instead of 1 because of the accumulation of harmful mutations or other reasons. A low reproductive rate is an initial reason promoting the accumulation of harmful mutations. Second, the critical reproductive rate of an asexual population $R*_1$ is higher than that of a sexual population $R*_2$. Mutation Specificity enables us to compare the advantages of sex and asexuality on the same figure. Third, a high reproductive rate can effectively prevent the accumulation of harmful mutations in a sexual or an asexual population.

Environmental conditions as available space, food supply, natural enemies and weather limit the reproductive capacity of a population. The mean reproductive rate has to reduce to 1 for a population which has reached to the carrying capacity even if there is not an enough accumulation of harmful mutations. The critical reproductive rate of an asexual population $R^*_1$ must be larger than 1. The critical reproductive rate of a sexual population $R^*_2$ is smaller than 1 except if the mutation rate is very high (Fig. 3). Therefore, an asexual population which has reached the carrying capacity becomes gradually extinct because of the reproductive rate lower than $R^*_1$ but a sexual population can effectively prevent extinction in the same case.

Reproductive rate plays a key role on the extinction of a population. A population must avoid a reproductive rate lower than $R^*$ to prevent extinction. Each population had to undergo reaching to the carrying capacity in evolutionary history because of the expansion. Therefore, sexual populations have survived and spread everywhere thanks to the $R^*$ smaller than 1. However, there are other reasonable reproductive approaches for organisms except for obligate sex. (i) Keep a high reproductive rate in comfortable conditions and stop producing in severe conditions. If it does this, a population need not reproduce sexually to prevent extinction. Two cases favour this effect: First, "lower" organisms, as most of protists, fungi, bacteria and some plants, have lower mutation rates thanks to small genomes so the $R^*$ reduces to close to 1. Thus, in a population having reached the carrying capacity some individuals can reproduce still with a rate larger than $R^*$ because of the diverse reproductive rates of individuals. Second, the extraordinary vitality of bdelloid rotifers (Gladyshev and Meselson 2008; Ricci 1998) enables them to survive and not to need reproduce in severe environment. (ii) Keep a high asexual reproductive rate in comfortable conditions and reproduce sexually in severe conditions. That is the facultative reproduction. This mechanism provides an alternative explanation for facultative organisms as yeast and aphid mentioned above. Under high stress conditions or in cold winter, yeasts and aphids cannot keep high reproductive rates because of not enough resources and the low survival rate of offspring so they reproduce provisionally sexually.

The influence of low fecundity on reproductive approach was discussed (Maynard Smith 1978; Williams 1975; Williams and Mitton 1973). Sexual reproduction is favourable when both fecundity and mortality are high because sex causes genetic variation. However, difficulties with low fecundities (as in humans) have been much greater (Hamilton *et al.* 1990). Fecundity is different from the reproductive rate in this study. That a high fecundity leads to a high mortality may be an approach conducing to select fitter individuals but may be only suitable for those species having small seeds costing less.

Reproductive rate affects initially the immediate evolutionary pressure from the accumulation of harmful mutations. Each population has a critical reproductive rate $R^*$. A population will be gradually extinct if the reproductive rate is lower than $R^*$. This can explain facultative, obligately asexual and sexual reproductions.

## Computer simulations

Equation 12, 13 and Eq. 11 reflect two kinds of relation between reproductive rate and harmful mutations. The important critical reproductive rate $R^*$ is derived from them. Equation 11 is unquestionable within the multiplicative mutation interactive model. Equation 6 and 10 are also obvious from the reasoning and the definition of Mutation Specificity. However, it is only an inference that the integrated harm *D* of mutations is related positively to Mutation Specificity *S* for individuals. Firstly, we will confirm the relation between *S* and *D* through simulations.

From section above, there are different evolutions in fitness with a sexual and an asexual population keeping a fixed size. Secondly, we verify whether the accumulation of harmful mutations causes the extinction of an asexual population and whether a sexual population can survive persistently, when reached the carrying capacity? Thirdly, whether an asexual population can prevent the accumulation of harmful mutations with a high reproductive rate?

For approximating a natural population, we consider both favourable and harmful mutation but ignore neutral mutation which has no effect on the fitness. Assume a harmful mutation has the effect *d* (*d* < 1) and a favourable mutation has the effect 1/*d*, so a favourable mutation can cancel out a harmful one. An individual with $m_1$ harmful and $m_2$ favourable mutations survives to reproduction with probability $d^{m_1}(1/d)^{m_2} = d^{m_1-m_2}$. New mutations for each individual arise with the frequency given by a Poisson distribution with mean *U*. A mutation is harmful with probability $P_d$. Assume *m* is the net number of harmful mutations, equal to the number of harmful mutations minus the number of favourable mutations ($m = m_1 - m_2$). Under the multiplicative mutation interactive model, the fitness *F* (the reproductive capacity) of an individual with $m_1$ harmful and $m_2$ favourable mutations is

$$F = F_0 \cdot d^{m_1-m_2} = F_0 \cdot d^m \tag{14}$$

where $F_0$ is the initial fitness.

A sexual or an asexual population, passing through many decades of generations, will accumulate enough mutations. We can calculate the integrated harm $D$ of mutations and Mutation Specificity $S$ of each individual in a population. When $g = 1$ with the $g$ of Eq. (3), there is not enough statistical significance to indicate that $D$ is positive related to $S$ (Fig. 4A). When $g = 5$, $D$ is clearly in direct proportion to $S$ (Fig. 4B). Therefore, the answer to the first question is that we should calculate Mutation Specificity with the parameter $g$ bigger than 1 and $D \propto S$. That also demonstrates Mutation Specificity is a perfect metric of the integrated harm of mutations within an individual.

The mean reproductive rate has to reduce to 1 for a population reaching the carrying capacity. That implies $N$ asexual or $N/2$ sexual mothers will reproduce only $N$ offspring. Assume an asexual population with $N$ adults, the fitness of adult $j$ is $F_j (1 \leq j \leq N)$. Adult $j$ produces stochastically an offspring with probability $F_j / \sum_{i=1}^{N} F_i$ and the total number of offspring is $N$. Each offspring inherits all mutations of her mother and gains new mutations which are harmful with probability $P_d$ and submit to a Poisson distribution with mean $U$. The net number of harmful mutations of an offspring is

$$m_{offspring} = m_{mother} + m_{new} \qquad (15)$$

The fitness of an offspring is

$$F_{offspring} = F_0 \cdot d^{m_{offspring}} \qquad (16)$$

With the alternation of generations, the fitness of each individual changes with the accumulation of mutations. A population is gradually extinct in numbered generations when the mean fitness reduces to lower than 1. A population persists continuously when the mean fitness of individuals keeps stabilized.

*N* adults can form stochastically *N/2* couples in a sexual population, where we do not consider the role of sexual selection. Each couple stochastically produces an offspring according to probability $(F_{male} + F_{female}) / \sum_{i=1}^{N} F_i$ and the total number of offspring is *N*. Each offspring inherits stochastically a mutation from parents with probability 0.5 and gains new mutations which are harmful with probability $P_d$ and submit to a Poisson distribution with mean *U*. The total number of harmful mutations of an offspring is

$$m_{offspring} = m_{inherit} + m_{new} \tag{17}$$

The fitness of an offspring can be obtained through Eq. 16.

Figure 5 shows that sexual and asexual populations evolve in different conditions. Assume the carrying capacity is 2000 (to prevent the randomness with a small population) and the probability that a mutation is harmful is 0.99 for populations. The results under four cases tell that each asexual population is undoubtedly extinct in a number of generations; in contrast, a sexual population can keep a stabilized mean fitness except if the initial fitness is low and mutations are strongly harmful. This has answered the second question. The results support significantly that a sexual population can effectively prevent the accumulation of harmful mutations but an asexual one cannot when reached the carrying capacity.

Assume a population reproduces in high mean reproductive rates (as *R* = 2, 3, 4 and 5) and the effect of new mutations might be large (*d* = 0.95) or small (*d* = 0.99).

Individuals reproduce quickly and the population expands under comfortable conditions. A disaster, in which most of individuals die, strikes and only few individuals (as 50) with higher fitness survive when the population exceeds the carrying capacity (as 2000). An asexual population evolves with the alternation of comfort and disaster. Figure 6 shows the results on two asexual populations with large or small new mutations under high reproductive rates. Two populations no longer decline in the mean fitness over time when the reproductive rate reaches to 5. The answer to the third question is that an asexual population with a high reproductive rate might prevent the accumulation of harmful mutations.

## Discussions

Some theories rely essentially on that sex increases genetic variation, as Red Queen hypothesis, Lottery model and Tangled Bank hypothesis. The "Fisher-Muller" hypothesis underlined that sex brings favourable mutations together. These theories reflect the long-term advantage of sex that helps a species to survive under a strong selection, such as a catastrophe. Muller's Ratchet shows that each individual of a population may decline in fitness because of the irreversible accumulation of harmful mutations. This mechanism, as a ratchet rotating in one direction only, brings individuals an immediate common pressure and may cause the extinction of an asexual population. These two kinds of theory, showing the long-term and short-term advantages of sex respectively, look like to have revealed the enigma of widespread sex.

However, our analysis shows that besides the accumulation of mutations determines the reproductive rate, the latter can affect on the former. The interaction of these two kinds of effect makes a critical reproductive rate $R^*$ for each lineage. A lineage even a population will be irreversibly extinct when the reproductive rate has reduced below $R^*$. In most cases, a sexual population has a critical reproductive rate below 1 and an asexual population has a reproductive rate above 1. The mean reproductive rate of a population may reduce to 1 because of reaching the carrying capacity or other reasons, so a sexual population with a lower $R^*$ is fitter than an asexual one. This mechanism can explain not only the maintenance of sex but also the existence of obligately asexual and facultative organisms. This theoretical advance might promote to understand clearly the short-term advantage of sex. Just like attaching a small new piece to a jigsaw puzzle, the enigma of sex appears as a more unabridged picture.

Mutation rate is positively related to $R^*$. The $R^*$ of a sexual lineage may be larger than 1 when the mutation rate is very high. If so, even a sexual population will be gradually extinct. Besides, the harm of mutations is diverse and mutations change their features with the environment, such as some neutral mutations become harmful or favourable when the environment changes. Mutation Specificity is suitable to approach the diversity of mutations from the definition. The influence of transforming mutations should be evaluated correctly in future studies.

A high Mutation Specificity implies mutations carrying polymorphisms. In most cases, the polymorphisms of mutations are equivalent to the genetic diversity of a population. Therefore, the genetic diversity implies a short-term harm rather than an advantage for a population. A moderate genetic diversity may be better for a population.

Researchers can only study a part not the full view of the evolution of sex through experiments. Many experiments are difficult to be finished because of the restriction of conditions. Therefore, experiments cannot replace the theoretical works although they are important. A theoretical description of the essence is very helpful for the deep understanding of evolution of sex, of course, more experiments are welcome.

An inference is that a low reproductive rate promotes the accumulation of harmful mutations so it is disadvantageous for a population even a sexual population. The rare favourable mutations will be swamped by many harmful mutations if we limit artificially and compulsively the reproductive rate of each family in a population. That is damaging for an evolving population from the viewpoint of natural selection. By this token, too rigorous and long-term an application of family planning might be inadvisable.

## Acknowledgment

We thank Lindi M. Wahl and Dietrich Stauffer for the helpful comments.

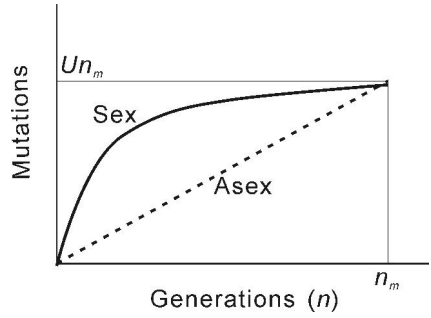

**Figure 1** The genetic variations of a sexual and an asexual population. The dashed line is the number of different mutations between two asexual individuals with a shared ancestor $n$ generations ago and the solid line is that of two sexual individuals. That implies sex causes genetic variation more quickly than asexuality. Notice, this figure does not imply that each population has the same $Un_m$. These lines reflect only a qualitative result based on Eq. 1 and 2 and the back mutations are ignored.

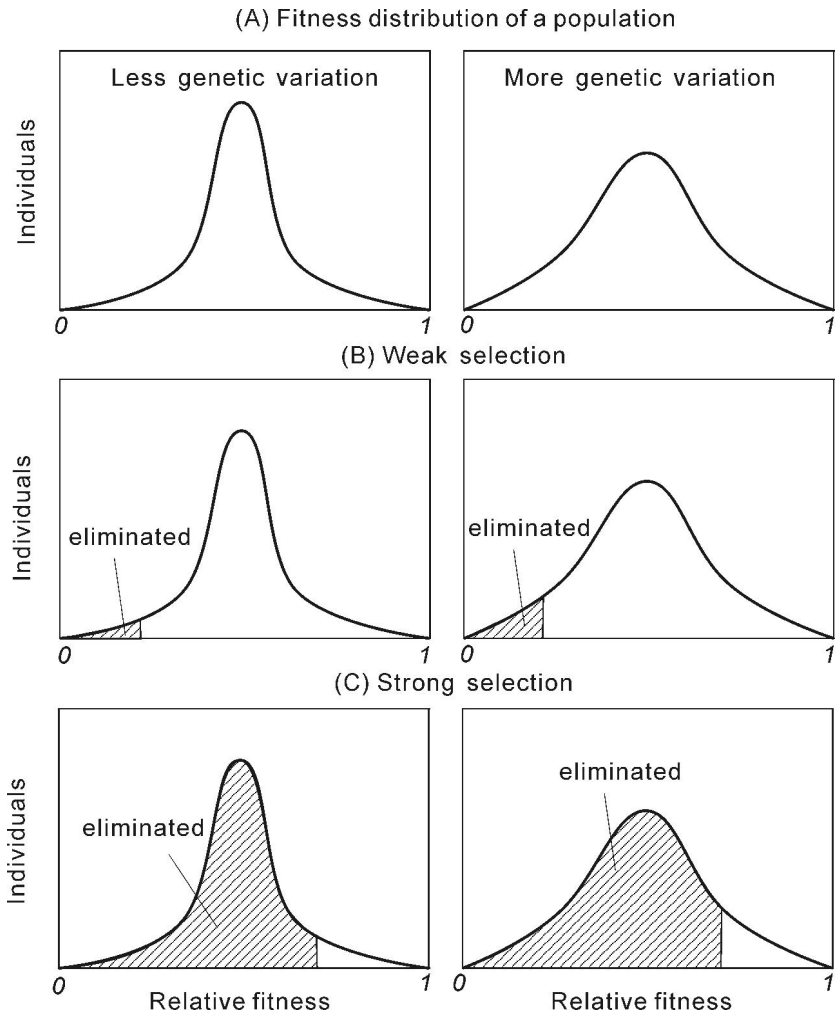

**Figure 2** The advantage of genetic variation (qualitative expression). (A) Individuals' relative fitness distributes more evenly in a population with more genetic variations than which with few genetic variations. (B) Under a weak selection, more individuals are eliminated (dashed area) in a population with more genetic variations than with few genetic variations. (C) Under a strong selection, more individuals survive in a population with more genetic variation than with few genetic variations.

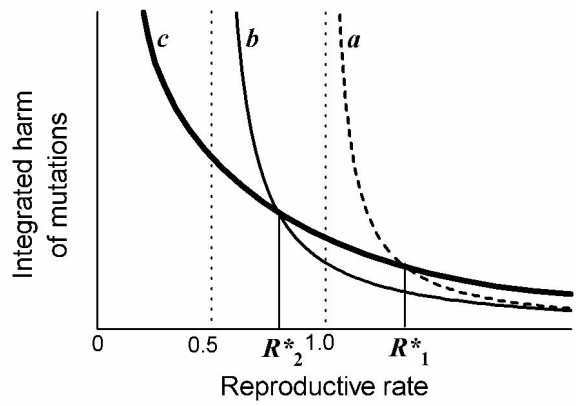

**Figure 3**  Integrated harm $D$ of mutations and reproductive rate $R$. Based on Mutation Specificity, reproductive rate affects the accumulation of harmful mutations (Eq. 12 and Eq. 13) in an asexual population (line $a$) and a sexual population (line $b$). Integrated harm of mutations restricts the fitness (reproductive capacity) (line $c$). $R^*_1$ and $R^*_2$ are the critical reproductive rates for an asexual and a sexual population.

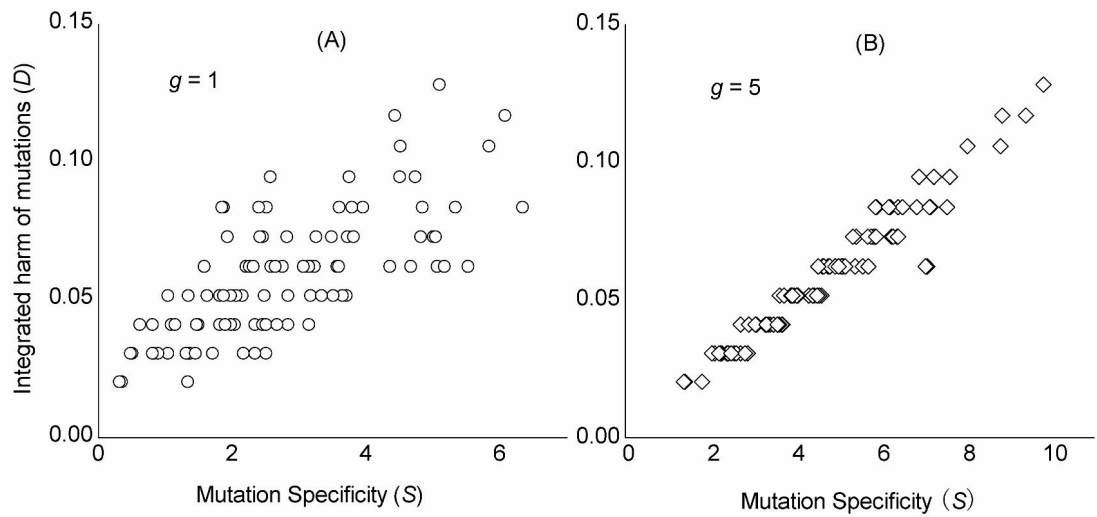

**Figure 4**　Relation between the integrated harm of mutations and Mutation Specificity. The scatter diagrams for (A) *g* = 1 and (B) *g* = 5 show that Mutation Specificity with a bigger *g* can measure better the accumulation of harmful mutations. The results are the same with a sexual and an asexual population.

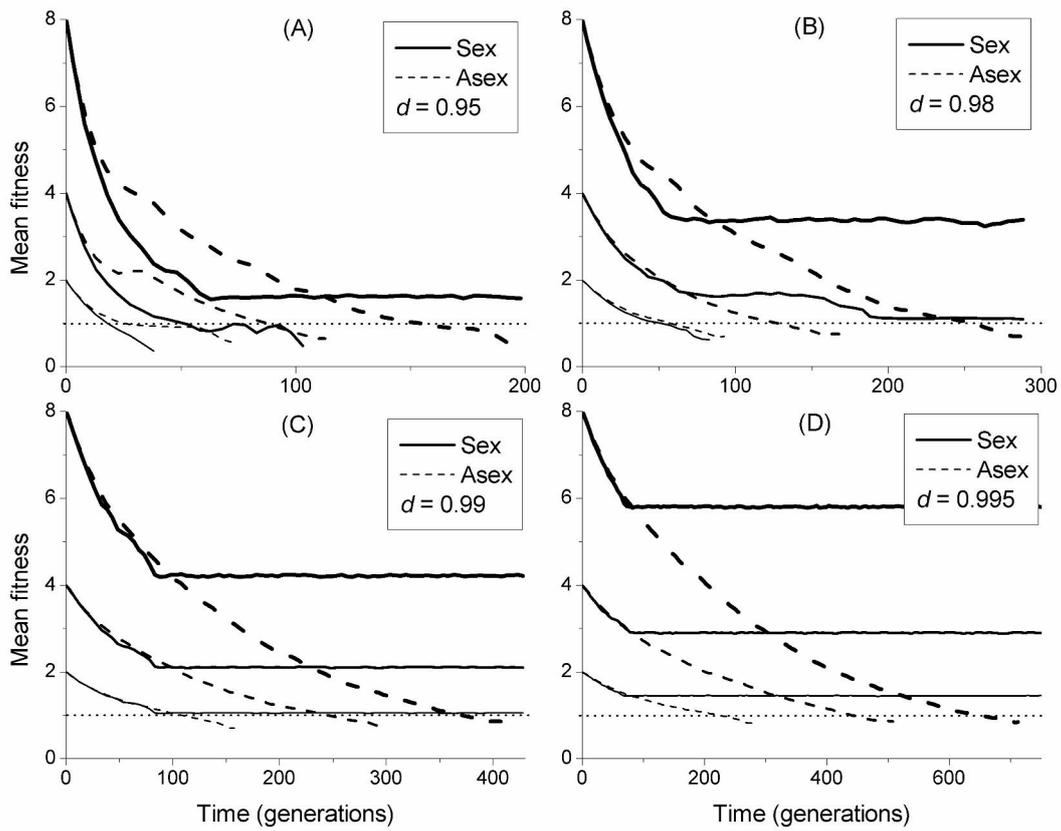

**Figure 5** Simulations under which sexual and asexual populations evolve under a low reproductive rate ($R$ = 1). Assume the number of offspring each generation is $N$ = 2000, the probability that a mutation is harmful is $P_d$ = 0.99 and $d$ is the effect of a mutation. New mutations arise in an individual with the frequency given by a Poisson distribution with mean 1. A mean fitness reduced to lower than 1 implies the extinction of a population.

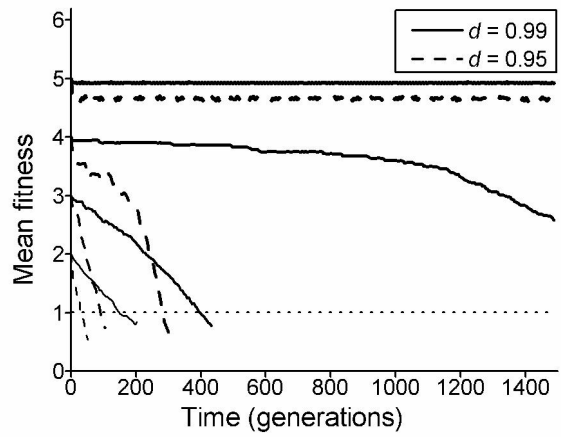

**Figure 6**  Simulations on asexual populations. Assume new mutations arise in an individual with the frequency given by a Poisson distribution with mean 1, the carrying capacity of a population is 2000 and 99% of mutations are harmful. Only fifty fitting individuals survive when population exceeds the carrying capacity. The populations with very unfavourable new mutations ($d$ = 0.05) and another with weakly unfavourable new mutations ($d$ = 0.01) are simulated under different reproductive rates ($R$ = 2, 3, 4 and 5, from bottom to top). Both populations keep stabilized fitness over time under a high reproductive rate ($R$ = 5).